\documentstyle[12pt,epsf,cite]{article}

\def\R{{\bf R}}
\def\C{{\bf C}}
\def\Z{{\bf Z}}
\def\uqsl{{\cal U}_q(sl(2,\C))}
\def\btuple#1#2#3#4#5#6{\left[
 \begin{array}{@{\,}ccc}
      #1 & #2 & #3 \\ #4 & #5 & #6
 \end{array}\right]}
\newcommand{\ei}[1]{e^{\epsilon_{#1}}_{i_{#1}}}
\newcommand{\eip}[1]{e^{\epsilon_{#1}}_{j_{#1}}}

\renewcommand{\Box}{\vbox{\hrule\hbox{\vrule\kern3pt
  \vbox{\kern6pt}\kern3pt\vrule}\hrule}}

\begingroup\makeatletter
\def\x#1#2#3#4#5#6#7\relax{\def\x{#1#2#3#4#5#6}}%
\expandafter\x\fmtname xxxxxx\relax %
\gdef\SetFigFont#1#2#3{%
  \ifnum #1<17\tiny\else \ifnum #1<20\small\else
  \ifnum #1<24\normalsize\else \ifnum #1<29\large\else
  \ifnum #1<34\Large\else \ifnum #1<41\LARGE\else
     \huge\fi\fi\fi\fi\fi\fi
  \csname #3\endcsname}%
\makeatother\endgroup

\def\abstract#1{\begin{center}{\large ABSTRACT}\end{center} \par #1}
\def\title#1{\begin{center}{\Large\bf {#1}}\end{center}}
\def\author#1{\begin{center}{\sc #1}\end{center}}
\def\address#1{\begin{center}{\it #1}\end{center}}

\begin{document}%%%%%%%%%%%%%%%%%%%%%%%%%%%%%%%%%%%%
\begin{titlepage}
\hspace*{\fill}
\vbox{
  \hbox{TIT/HEP-292}
% \hbox{UT-KOMABA/???}
  \hbox{hep-th/9507146}
  \hbox{July 1995}}
\vspace*{\fill}
\begin{center}
  \Large\bf
\mbox{A connection
  between lattice and surgery constructions}
  of three-dimensional topological field theories
\end{center}

\vskip 1cm
\author{
  Masako ASANO \footnote{E-mail address: \tt maa@th.phys.titech.ac.jp}}
\address{
  Department of Physics,
  Tokyo Institute of Technology, \\
  Oh-Okayama, Meguro,  Tokyo 152, Japan}
\begin{center}
  and
\end{center}
\author{
  Saburo HIGUCHI
  \footnote{E-mail address: \tt hig@rice.c.u-tokyo.ac.jp}}
\address{
  Department of Pure and Applied Sciences,  College of Arts and Sciences\\
  University of Tokyo\\
  Komaba, Meguro, Tokyo 153, Japan
}
\vspace{\fill}
\abstract{
  We study the relation between
  lattice construction and surgery construction
  of three-dimensional topological field theories.
  We show that a class of
  the Chung-Fukuma-Shapere theory on the lattice
  has  representation theoretic reformulation
  which is closely related to the Altschuler-Coste theory constructed
  by surgery.
  There is a similar relation
  between the Turaev-Viro theory
  and the Reshetikhin-Turaev theory.
  } \vspace*{\fill}
\end{titlepage}
%%%%%%%%%%%%%%%%%%%%%%%%%%%%%%%%%%%%%%%%%%%%%%%%%%%%%%%%%%%%%%%%%%%%%%
Various three-dimensional topological field theories
which satisfy Atiyah's axiom \cite{a}
have been constructed by now~\cite{dw,tv,rt}. %\cite{ku}.
Relations among them are, however, still not clear.
It is important to investigate them more
with a view to systematic classifications of three-dimensional
topological field theories \footnote{
Note that two-dimensional unitary topological field theories are
classified completely~\cite{dj}.}.
In the following, we study several topological field theories
and
establish a connection among them.

There are two principal methods to construct
three-dimensional topological field theories
with mathematical rigor.
One is the lattice or `state-sum' method in which
we represent a manifold by a
simplicial complex
and consider a statistical model on it.
The other method employs the surgery representation of three-manifolds.
One can construct arbitrary closed three-manifold $M_{(L,f)}$
by a surgery of $S^3$ along a framed link $(L,f)$ in it.
Therefore a class of framed link invariants, actually those which are
invariant under Kirby moves,
give rise to invariants of three-manifolds.
Sometimes this construction is lifted to that of a topological field theory
whose partition function is the three-manifold invariant.

Though these two methods differ a lot in nature,
surprisingly topological field theories
defined by
the two
methods sometimes reveal close relationship.
The most famous example is the relation between
the lattice theory by Turaev and Viro
\cite{tv} (TV) and the Reshetikhin-Turaev (or
Chern-Simons~\cite{wi}) theory~\cite{rt} (RT)
defined by surgery:
$Z^{TV}=|Z^{RT}|^2$~\cite{tu}.
There is also a suggested relation~\cite{ac,ah} between
the Dijkgraaf-Witten~\cite{dw} or
Chung-Fukuma-Shapere~\cite{cfs} theory on the lattice,
and the Altschuler-Coste theory~\cite{ac} constructed using the
surgery representation.
It is desirable to have a more general statement
which relates  the lattice construction  and the surgery construction.

In this note, we study whether a functor used in the surgery
construction of topological field theories can  induce a
topological field theory on lattice.
We take the Altschuler-Coste functor $F^{AC}$ employed in the surgery
construction in ref.\cite{ac} as an example and show that it is
possible.

We are motivated by a remark in ref.\cite{bou2} on the relation
between the Turaev-Viro theory and the Reshetikhin-Turaev theory.
In ref.\cite{bou2}, an attempt was made
to rewrite the Turaev-Viro invariant in the form of the partition
function of a `three-dimensional $q$-deformed lattice gauge theory,'
defined making use of the functor $F^{RT}_{\uqsl}.$
It is a functor from the category of
colored ribbon graphs to the category of representations of a
ribbon Hopf algebra
and was originally used to define the Reshetikhin-Turaev theory.

We first review the properties of the functor $F^{AC}$
and then show how a three-manifold invariant on the lattice is derived
from it and that it is equivalent with the partition function of
the Chung-Fukuma-Shapere theory.
We see, however, that the Turaev-Viro theory is not reproduced from
the Reshetikhin-Turaev functor
except in the limit $q \rightarrow 1$
if we apply the same prescription in this case.
\bigskip

%%%%%%%%%%%%%%%%%%%%%%%%%%%%%%%%%%%%%%%%%%%%%%%%%%%%%%%%%%%%%%%%%%%%%%
The functor $F^{AC}$ is a functor from the category of
colored ribbon graphs to the category of representations of
the quasi-Hopf algebra $D^{\omega}(G)$.
The algebra $D^\omega(G)$ is defined for a finite group $G$ and its 3-cocycle
$\omega : G\times G\times G\rightarrow {\rm U}(1)$, which
satisfies
\begin{equation}
\omega(g,x,y)\omega(gx,y,z)^{-1}\omega(g,xy,z)
\omega(g,x,yz)^{-1}\omega(x,y,z)=1 .
\end{equation}
The condition $\omega(x,y,z)=1$ should also be met
if at least one of the arguments $x,y,z$ is equal to the unit element $e\in G$.
The bialgebra $D^{\omega}(G)$
is spanned by a formal basis $\{\chi_{g,x}|g,x\in G\}$ as a $\C$-module.
Multiplication, unit, comultiplication and
counit on $D^{\omega}(G)$
is defined with respect to the basis $\{\chi_{g,x}\}$
as follows :
\begin{eqnarray}
  \lefteqn{\chi_{g,x} \cdot \chi_{h,y} =\delta_{g,xhx^{-1}}
    \theta_g(x,y)\;\chi_{g,xy}}
         & &\hspace*{8cm}
         {\rm (multiplication)}\\
  \lefteqn{ u(1\in {\bf C}) = \chi_{1,e} \equiv \sum_{g\in G}\chi_{g,e}}
                 & & \hspace*{8cm} {\rm (unit)}\\
  \lefteqn{\Delta (\chi_{g, h})=\sum_{xy=g}\gamma_h(x,y)\;\chi_{x,h}\otimes
       \chi_{ y,h}}
           & & \hspace*{8cm}  {\rm (comultiplication )}\\
  \lefteqn{\epsilon (\chi_{ g,e})=\delta_{g,h}\in\C}
           & & \hspace*{8cm} {\rm (counit)}
\end{eqnarray}
where
\begin{eqnarray}
\theta_g(x,y)&\equiv& \omega(g,x,y)\;\omega(x,y,(xy)^{-1}gxy)\;
   \omega(x,x^{-1}gx,y)^{-1},\label{theta}\\
\gamma_x(g,h) &\equiv& \omega(g,h,x)\;\omega(x,x^{-1}gx,x^{-1}hx)\;
   \omega(g,x,x^{-1}hx)^{-1}  \label{gamma}.
\end{eqnarray}
Furnished with the antipode $S$, the $R$-matrix, and an element $\phi $,
the bialgebra $D^{\omega}(G)$ becomes quasi-triangular quasi-Hopf
algebra :
\begin{eqnarray}
  S(\chi_{g,h}) & = & \theta_{g^{-1}}(h,h^{-1})^{-1}
  \gamma_h(g,g^{-1})^{-1} \;\chi_{h^{-1}g^{-1}h,h^{-1}}, \\
  R & =&\sum _{g,h\in G}\;\chi_{g,e}\otimes\chi_{h,g},\\
  \phi &=&\sum_{g,h,k\in G}\omega(g,h,k)^{-1}\chi_{g,e}
  \otimes \chi_{h,e}\otimes\chi_{k,e}.
\end{eqnarray}

Now to define $F^{AC}$, we explain the notion of colored ribbon graph $\Gamma$.
We provide a ribbon and an annulus
which are a square $[0,1]\times [0,1]$
and a cylinder $[0,1]\times S^1$
embedded in $\R^2 \times [0,1]$.
For an annulus $C$,
the linking number $lk(C^+,C^-)$ is called the framing of the annulus $C$
where $C^+$ and $C^-$ denote the images of the circles
$0\times S^1$ and  $1\times S^1$, respectively.

Each ribbon (or annulus) is equipped with an arrow along
the image of $(1/2)\times [0,1]$ (or $(1/2) \times S^1$ ) and
we assume  the right side and the wrong side
are defined on it.
Given two non-negative integers $k$ and $l$,
a ribbon $(k,l)$-graph $\Gamma$ is defined as a
disjoint union of finite number of ribbons and annuli.
Two ends $[0,1]\times 0$ and $[0,1]\times 1$ of the ribbons are
mapped onto
\begin{eqnarray*}
\Gamma\cap (\R^2\times 0)
&=& \cup_{i=1}^k\left(\left\{\,0\times [i-1/4, i+1/4]\times 0\right\}\right),\\
\Gamma\cap (\R^2\times 1)
&=&\cup_{i=1}^l \left(\left\{\,0\times [i-1/4, i+1/4]\times 1 \right\}\right).
\end{eqnarray*}
Near the two ends of each ribbon, the right side is faced to the plus
direction of the $x$-axis.

\begin{figure}
    \begin{center}
      \leavevmode
       \epsfysize = 5cm
       \epsfbox{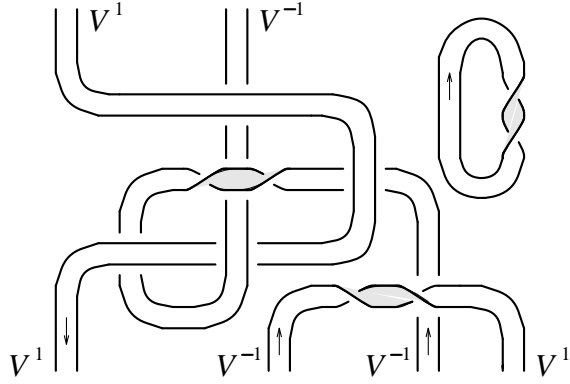}
    \end{center}
    \caption{A ribbon $(4,2)$-graph colored by, for example,
       $w_4^{(b)}=(V^{1}\Box ((V^{-1}\Box V^{-1})\Box V^{1}))$ and
       $w_2^{(t)}=(V^{1}\Box V^{-1})$.}
    \label{cgraph}
  \end{figure}

Furthermore, we introduce `ribbon graphs colored by regular representations
of $D^{\omega}(G)$,' or `regular $c$-graphs.'
A regular $c$-graph is a
ribbon $(k,l)$-graph together with
a choice of formal symbols, or {\sl words},
$w_k^{(b)}$ and $w_l^{(t)}$ associated
with the bottom end and the top end of $\Gamma$.
An example of $w_7$ is
\begin{equation}
((V^{\epsilon_1} \Box (( V^{\epsilon_2} \Box V^{\epsilon_3}) \Box
V^{\epsilon_4})) \Box
(V^{\epsilon_5} \Box (V^{\epsilon_6} \Box V^{\epsilon_7}))),
\end{equation}
where $\Box$ is a formal non-associative binary operator.
The symbols $\epsilon_i$ should take values $+1$ or $-1$
when the $i$-th ribbon is directed downward or upward, respectively.
The symbol $V^1$ stands for the $|G|^2$-dimensional
$D^{\omega}(G)$-module acted by  the regular representation of $D^{\omega}(G)$
and $V^{-1}$  its dual.
In general the word associated with the top (bottom) of a $c$-graph is
not unique because there are several ways of putting  parentheses (i.e. the
order of the operation of $\Box$).
An example of regular $c$-graphs is depicted in Fig.~\ref{cgraph}.
%%%%%%%%%%%%%%%%%%
  %%%%%%%%%%%%%%%%%%

The functor $F^{AC}$ maps a regular colored ribbon $(k,l)$-graph $\Gamma$ to
a linear map
\begin{equation}
  F^{AC}(\Gamma) : V^{\epsilon_1}\otimes V^{\epsilon_2}\otimes \cdots
  \otimes V^{\epsilon_k}\;\longrightarrow\;
  V^{\epsilon_1}\otimes V^{\epsilon_2}\otimes\cdots
  \otimes V^{\epsilon_l} .
\end{equation}
The explicit form of the map $F^{AC}(\Gamma)$
is determined by the composition of $F^{AC}(\Gamma')$ for elementary graphs
$\Gamma'$ given as follows
: \footnote{A single line
represents
a ribbon whose right side is faced to us.
}
%%%%%%%%%%%%%%%%%%%%%%
\begin{eqnarray}
   F^{AC}\Bigg(\;
   \begin{picture}(50,30)
     \put(25,20){\line(0,-1){34}}
     \put(22,25){$V^{\epsilon}$}
     \put(22,-25){$V^{\epsilon}$}
   \end{picture}
   \Bigg)& =& \mbox{id}_{V^\epsilon}\; :\; V^\epsilon\rightarrow V^\epsilon
   \rule[-1cm]{0cm}{2.2cm}
   \label{op:id}\\
   F^{AC}\Bigg(\;
   \begin{picture}(50,30)
     \put(23,10){\oval(40,40)[br]}
     \put(27,10){\oval(40,40)[bl]}
     \put(23,-10){\vector(1,0){4}}
     \put(5,15){$V^{1}\Box\; V^{-1}$}
   \end{picture}
   \Bigg)\;(1)\; & =& \sum_{g,h}\; \omega(g,g^{-1},g)
               \chi_{g,h}\otimes\psi_{g,h}
     \; :\: {\bf C} \rightarrow V\otimes V^{-1}
   \rule[-1cm]{0cm}{2.2cm} \label{op:cupright}\\
   F^{AC}\Bigg(\;
   \begin{picture}(50,30)
     \put(23,10){\oval(40,40)[br]}
     \put(27,10){\oval(40,40)[bl]}
     \put(27,-10){\vector(-1,0){4}}
     \put(5,15){$V^{-1}\Box V^{1}$}
   \end{picture}
   \Bigg)\;(1)\; & =& \sum_{g,h}\; \psi_{g,h}\otimes\chi_{g,h}
     \; :\: {\bf C} \rightarrow V^{-1}\otimes V
   \rule[-1cm]{0cm}{2.2cm}\label{op:cupleft}\\
   F^{AC}\Bigg(\;
   \begin{picture}(50,30)
     \put(23,-2){\oval(40,40)[tr]}
     \put(27,-2){\oval(40,40)[tl]}
     \put(23,18){\vector(1,0){4}}
     \put(3,-15){$V^{-1}\Box V^{1}$}
   \end{picture}
   \Bigg)\;(\psi_{g,h}\otimes\chi_{x,y})
   \; &= &\delta_{g,x}\,\delta_{h,y}
     \; :\: V^{-1}\otimes V \rightarrow {\bf C}
   \rule[-1cm]{0cm}{2.2cm}\label{op:capright}\\
   F^{AC}\Bigg(\;
   \begin{picture}(50,30)
     \put(23,-2){\oval(40,40)[tr]}
     \put(27,-2){\oval(40,40)[tl]}
     \put(27,18){\vector(-1,0){4}}
     \put(3,-15){$V^{1}\Box\; V^{-1}$}
   \end{picture}
   \Bigg)\;(\chi_{g,h}\otimes\psi_{x,y})
   \; & =& \omega(g^{-1}, g, g^{-1})\,\delta_{g,x}\,\delta_{h,y}
     \; :\: V\otimes V^{-1} \rightarrow {\bf C}
   \rule[-1cm]{0cm}{2.2cm} \label{op:capleft}\\
  F^{AC}\Bigg(\;
 \begin{picture}(50,30)
    \put(8,-10){\line(1,1){25}}
    \put(33,-10){\line(-1,1){11}}
    \put(8,15){\line(1,-1){11}}
    \put(0,-23){$V^{\epsilon_1}\Box\;\, V^{\epsilon_2}$}
    \put(0,20){$V^{\epsilon_2}\Box\;\, V^{\epsilon_1}$}
 \end{picture}
   \Bigg)\;
  &=& P_{12}\circ (\pi^{\epsilon_1}\otimes \pi^{\epsilon_2})\; (R)
         \nonumber\\
  & &  \hspace*{1cm}\; :\;
      V^{\epsilon_1}\otimes V^{\epsilon_2} \rightarrow
      V^{\epsilon_2}\otimes V^{\epsilon_1}
   \rule[-1cm]{0cm}{1.2cm} \\
  F^{AC}\Bigg(\;
 \begin{picture}(50,30)
    \put(8,-10){\line(1,1){11}}
    \put(33,-10){\line(-1,1){25}}
    \put(33,15){\line(-1,-1){11}}
    \put(0,-23){$V^{\epsilon_2}\Box\;\, V^{\epsilon_1}$}
    \put(0,20){$V^{\epsilon_1}\Box\;\, V^{\epsilon_2}$}
 \end{picture}
   \Bigg)\;
  &=& (\pi^{\epsilon_1}\otimes \pi^{\epsilon_2})\; (R^{-1})\circ P_{21}
          \nonumber\\
  & &  \hspace*{1cm}\; :\;
    V^{\epsilon_2}\otimes V^{\epsilon_1} \rightarrow
      V^{\epsilon_1}\otimes V^{\epsilon_2}
   \rule[-1cm]{0cm}{1.2cm} \\
  F^{AC}\Bigg(\;
 \begin{picture}(80,30)
    \put(8,-13){\line(0,1){34}}
    \put(35,-13){\line(0,1){34}}
    \put(62,-13){\line(0,1){34}}
    \put(0,-25){$V^{\epsilon_1}\Box (V^{\epsilon_2}\Box V^{\epsilon_3})$}
    \put(0,25){$(V^{\epsilon_1}\Box V^{\epsilon_2})\Box V^{\epsilon_3}$}
  \end{picture}
   \Bigg)\;
 &  = &(\pi_1^{\epsilon_1}\otimes\pi_2^{\epsilon_2}\otimes\pi_3^{\epsilon_3})
   \;(\phi) \nonumber\\
   & &  \; :\;
   V^{\epsilon_1}\otimes V^{\epsilon_2}\otimes V^{\epsilon_3}
   \rightarrow
      V^{\epsilon_1}\otimes V^{\epsilon_2}\otimes V^{\epsilon_3}
  \label{op:phi}
\end{eqnarray}
%%%%%%%%%%%%%%%%%%%%%
where $P_{12}$ denotes the  permutation operator :
$$P_{12}: a_1\otimes b_2\mapsto b_2\otimes a_1 :
\qquad
V^{\epsilon_1}\otimes V^{\epsilon_2} \rightarrow
      V^{\epsilon_2}\otimes V^{\epsilon_1}.$$
There appear  the  regular representation $\pi^1(=\pi)$
and its dual $\pi^{-1}(=\pi^\ast)$.
Explicitly, they are
 \begin{eqnarray}
   \pi(a)\, x &\equiv & a\cdot x \\
   \pi^{\ast}(a)\, x^{\ast}&\equiv & x^\ast\left( \pi\circ S(a)\right)
 \end{eqnarray}
where $a,x\in D^{\omega}(G)$ and $x^\ast\in D^{\omega}(G)^*$.

\bigskip
We proceed to defining a three-manifold invariant
using a lattice and  the functor $F^{AC}$.
Our basic strategy is as follows.
We provide a triangulation $T$ of a closed three-manifold $M$
and associate a framed link $L_T$ in $M$ with $T$.
Then we try to operate the functor $F^{AC}$ on $L_T$ `interpreted' as a
colored $(0,0)$-graph $\Gamma_T$.

We associate a framed link $L_T$ with a triangulation $T$ in the following way:
\begin{enumerate}
 \item
  With each $2$-simplex $f_i$ in $T$, we associate a trivial knot $C_i$ along
  the boundary $\partial f_i$, which yields a trivial link with $n_2$
  components (We denote by $n_i$ the number of $i$-simplices in $T$).
\item
  To the link $\cup_{i=1}^{n_2}C_i$,
  we add  components $C_{n_2+j} \; (j=1,\ldots,n_1)$ so as to encircle
  each bundle of segments of knots
  corresponding to each $1$-simplex $j$ (Fig.\ref{lin}~(b)).
  We obtain a link $L_T$  with $n_1 +n_2$ components  altogether:
  $L_T=\cup_{i=1}^{n_2+n_1}C_i$.
  %%%%%%%%%%%%%%%%%%
  \begin{figure}[bht]
    \begin{center}
      \leavevmode
       \epsfxsize = 16cm
       \epsfbox{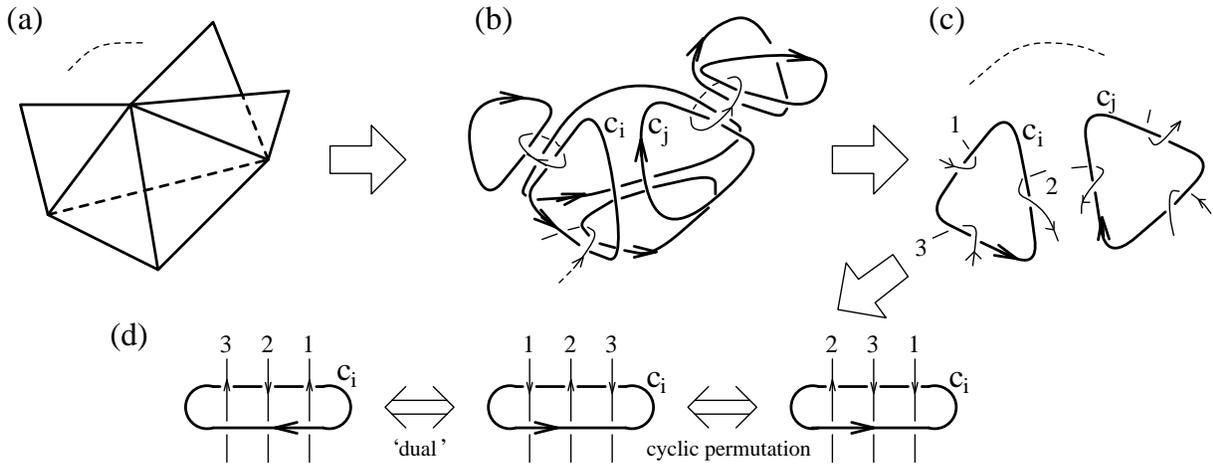}
    \end{center}
    \caption{The link representation of $T$ :
     (a)~A part of $T$. (b)~A part of $L_T$ with arrows.
     (c)~The decomposition into building blocks. Two building
     blocks are drawn.
     (d)~Interpretation as ribbon $(3,3)$-graphs.}
    \label{lin}
  \end{figure}
  %%%%%%%%%%%%%%%%%%
\item
  We thicken each component in $L_T$ to be an annulus of $0$
  framing~\footnote{
This makes sense because each link component is contained in a 3-ball in $M$.
}
     and   associate an arrow along each annulus arbitrarily to have a
     framed link.
\end{enumerate}

Let us pause for a moment here to explain
why we construct framed links in this way.
Eventually we would hope
to interpret $L_T$ as a ribbon $(0,0)$-graph $\Gamma_T$ and
to prove that $F^{AC}(\Gamma_T)$ is
invariant under the Alexander moves \cite{al} of $T$.
Since we have a map from triangulations to framed links, we can
determine a set of local modifications of links  that corresponds to
Alexander moves.
In fact, such a set of local modification is
generated by the Kirby moves, under which $F^{AC}(\Gamma_T)$ is invariant.
Thus we would expect that the correspondence between triangulations
and framed links is a key relation which connects lattice and surgery
constructions of topological field theories.

However here we face the fact that $F^{AC}(\Gamma_T)$ is not
well-defined.
This is because $\Gamma_T$ should be in $\R^3$ while $L_T$ is contained in
the manifold $M$.
So our task is not so straightforward.

Nevertheless, it is possible to generalize the action of
$F^{AC}$ to yield a number for the framed link obtained above.
We first decompose the framed link $L_T$ into
$n_2$ building blocks each of which corresponds to a $2$-simplex as
depicted in Fig.\ref{lin}~(c).
Then we decorate them to make them  $(3,3)$ $c$-graphs (Fig.\ref{lin}(d)).
The framings are derived from those of framed links.
Arrows are already associated with the components.
We decide which end is the top of the graph arbitrarily.
This fixes the order of the three vertical ribbons up to a cyclic
permutation.
We also associate words with the bottom end and the top end of a
$(3,3)$-graph arbitrarily under the restriction that the $\epsilon$'s
match the direction of the arrows.
It is possible to associate two different words with the
bottom (top) end of the graph.
Namely it  can be
$ (V^{\epsilon_1}\Box V^{\epsilon_2})\Box V^{\epsilon_3}$ or
$ V^{\epsilon_1}\Box (V^{\epsilon_2}\Box V^{\epsilon_3})$.
If we  set the position of two parentheses
in $w_3{}^{(t)}$ and $w_3{^{(b)}}$
to be identical, we have only to deal with two kinds of $c$-graphs:
 \begin{equation}
    \Gamma_{({\epsilon_1},{\epsilon_2}),{\epsilon_3}}
    \equiv
\setlength{\unitlength}{0.003375in}%
\begin{picture}(272,150)(-12,750)
\thinlines
\put( 45,750){\oval( 60, 60)[tl]}
\put( 45,750){\oval( 60, 60)[bl]}
\put(255,750){\oval( 60, 60)[br]}
\put(255,750){\oval( 60, 60)[tr]}
\put( 45,765){\oval( 60, 60)[tl]}
\put( 45,765){\oval( 60, 60)[bl]}
\put(255,765){\oval( 60, 60)[br]}
\put(255,765){\oval( 60, 60)[tr]}
\put( 45,718){\line( 1, 0){210}}
\put( 45,733){\line( 1, 0){210}}
\put(225,711){\line( 0,-1){ 36}}
\put(240,711){\line( 0,-1){ 36}}
\put( 60,711){\line( 0,-1){ 36}}
\put( 75,711){\line( 0,-1){ 36}}
\put( 60,744){\line( 0, 1){ 81}}
\put( 75,825){\line( 0,-1){ 81}}
\put(225,744){\line( 0, 1){ 81}}
\put(240,825){\line( 0,-1){ 81}}
\put(144,744){\line( 0, 1){ 81}}
\put(159,825){\line( 0,-1){ 81}}
\put(144,711){\line( 0,-1){ 36}}
\put(159,711){\line( 0,-1){ 36}}
\put( 45,795){\line( 0,-1){ 15}}
\put( 39,795){\line( 0,-1){ 15}}
\put( 33,792){\line( 0,-1){ 15}}
\put( 27,789){\line( 0,-1){ 15}}
\put( 21,783){\line( 0,-1){ 15}}
\put( 15,768){\line( 0,-1){ 15}}
\put(255,795){\line( 0,-1){ 15}}
\put(261,795){\line( 0,-1){ 15}}
\put(267,792){\line( 0,-1){ 15}}
\put(273,789){\line( 0,-1){ 15}}
\put(279,783){\line( 0,-1){ 15}}
\put(285,768){\line( 0,-1){ 15}}
\put( 84,795){\line( 1, 0){ 51}}
\put( 84,780){\line( 1, 0){ 51}}
\put(168,795){\line( 1, 0){ 48}}
\put(168,780){\line( 1, 0){ 48}}
\put( 90,795){\line( 0,-1){ 15}}
\put( 96,795){\line( 0,-1){ 15}}
\put(102,795){\line( 0,-1){ 15}}
\put(108,795){\line( 0,-1){ 15}}
\put(114,795){\line( 0,-1){ 15}}
\put(120,795){\line( 0,-1){ 15}}
\put(126,795){\line( 0,-1){ 15}}
\put(132,795){\line( 0,-1){ 15}}
\put(138,795){\line( 0,-1){ 15}}
\put(168,795){\line( 0,-1){ 15}}
\put(174,795){\line( 0,-1){ 15}}
\put(180,795){\line( 0,-1){ 15}}
\put(186,795){\line( 0,-1){ 15}}
\put(192,795){\line( 0,-1){ 15}}
\put(198,795){\line( 0,-1){ 15}}
\put(204,795){\line( 0,-1){ 15}}
\put(210,795){\line( 0,-1){ 15}}
\put(216,795){\line( 0,-1){ 15}}
\put( 84,795){\line( 0,-1){ 15}}
     \put(6,638){$(V^{\epsilon_1}\Box
         V^{\epsilon_2})\Box V^{\epsilon_3}$}
     \put(6,850){$(V^{\epsilon_1}\Box V^{\epsilon_2})\Box
        V^{\epsilon_3}$}
 \end{picture}%\;\;\;\;\;\;\;\;\; .
 \rule[-2cm]{0mm}{3.5cm}
 \end{equation}
and
 \begin{equation}
    \Gamma_{{\epsilon_1},({\epsilon_2},{\epsilon_3})}
    \equiv
\setlength{\unitlength}{0.003375in}%
\begin{picture}(272,150)(-12,750)
\thinlines
\put( 45,750){\oval( 60, 60)[tl]}
\put( 45,750){\oval( 60, 60)[bl]}
\put(255,750){\oval( 60, 60)[br]}
\put(255,750){\oval( 60, 60)[tr]}
\put( 45,765){\oval( 60, 60)[tl]}
\put( 45,765){\oval( 60, 60)[bl]}
\put(255,765){\oval( 60, 60)[br]}
\put(255,765){\oval( 60, 60)[tr]}
\put( 45,718){\line( 1, 0){210}}
\put( 45,733){\line( 1, 0){210}}
\put(225,711){\line( 0,-1){ 36}}
\put(240,711){\line( 0,-1){ 36}}
\put( 60,711){\line( 0,-1){ 36}}
\put( 75,711){\line( 0,-1){ 36}}
\put( 60,744){\line( 0, 1){ 81}}
\put( 75,825){\line( 0,-1){ 81}}
\put(225,744){\line( 0, 1){ 81}}
\put(240,825){\line( 0,-1){ 81}}
\put(144,744){\line( 0, 1){ 81}}
\put(159,825){\line( 0,-1){ 81}}
\put(144,711){\line( 0,-1){ 36}}
\put(159,711){\line( 0,-1){ 36}}
\put( 45,795){\line( 0,-1){ 15}}
\put( 39,795){\line( 0,-1){ 15}}
\put( 33,792){\line( 0,-1){ 15}}
\put( 27,789){\line( 0,-1){ 15}}
\put( 21,783){\line( 0,-1){ 15}}
\put( 15,768){\line( 0,-1){ 15}}
\put(255,795){\line( 0,-1){ 15}}
\put(261,795){\line( 0,-1){ 15}}
\put(267,792){\line( 0,-1){ 15}}
\put(273,789){\line( 0,-1){ 15}}
\put(279,783){\line( 0,-1){ 15}}
\put(285,768){\line( 0,-1){ 15}}
\put( 84,795){\line( 1, 0){ 51}}
\put( 84,780){\line( 1, 0){ 51}}
\put(168,795){\line( 1, 0){ 48}}
\put(168,780){\line( 1, 0){ 48}}
\put( 90,795){\line( 0,-1){ 15}}
\put( 96,795){\line( 0,-1){ 15}}
\put(102,795){\line( 0,-1){ 15}}
\put(108,795){\line( 0,-1){ 15}}
\put(114,795){\line( 0,-1){ 15}}
\put(120,795){\line( 0,-1){ 15}}
\put(126,795){\line( 0,-1){ 15}}
\put(132,795){\line( 0,-1){ 15}}
\put(138,795){\line( 0,-1){ 15}}
\put(168,795){\line( 0,-1){ 15}}
\put(174,795){\line( 0,-1){ 15}}
\put(180,795){\line( 0,-1){ 15}}
\put(186,795){\line( 0,-1){ 15}}
\put(192,795){\line( 0,-1){ 15}}
\put(198,795){\line( 0,-1){ 15}}
\put(204,795){\line( 0,-1){ 15}}
\put(210,795){\line( 0,-1){ 15}}
\put(216,795){\line( 0,-1){ 15}}
\put( 84,795){\line( 0,-1){ 15}}
     \put(6,638){$V^{\epsilon_1}\Box
         (V^{\epsilon_2}\Box V^{\epsilon_3})$}
     \put(6,850){$V^{\epsilon_1}\Box (V^{\epsilon_2}\Box
        V^{\epsilon_3})$}
 \end{picture}\;\;\;\;\;\;\;\;\; .
 \rule[-2cm]{0mm}{3.5cm}
 \end{equation}
Here $V^{\epsilon_i}$ denotes the vector space
$V^{\epsilon_1}$ associated with a ribbon $i$.
The relation between the two maps corresponding to these two
$c$-graphs is determined by (\ref{op:phi}) as
 \begin{eqnarray}
 \lefteqn{  F^{AC}
 (\Gamma_{{\epsilon_1},({\epsilon_2},{\epsilon_3})}) }
 \nonumber\\
   &=&
   %\pi^{\epsilon_1 \epsilon_2 \epsilon_3 }(\phi^{-1})
   (\pi_1^{\epsilon_1}\otimes\pi_2^{\epsilon_2 }
      \otimes\pi_3^{\epsilon_3 })(\phi^{-1})
   \circ
   F^{AC}
 (\Gamma_{({\epsilon_1},{\epsilon_2}),{\epsilon_3}})
   \circ
   %\pi^{\epsilon_1 \epsilon_2 \epsilon_3 }(\phi)\\
   (\pi_1^{\epsilon_1}\otimes\pi_2^{\epsilon_2 }
      \otimes\pi_3^{\epsilon_3 })(\phi)\\
   & & \hspace*{3cm}
      \, : \, V^{\epsilon_1}\otimes V^{\epsilon_2}\otimes
      V^{\epsilon_3} \longrightarrow
      V^{\epsilon_1}\otimes V^{\epsilon_2}\otimes V^{\epsilon_3} \nonumber.
 \end{eqnarray}
Now, let us demand that
\begin{equation}
  F^{AC}\left(\Gamma_{{\epsilon_1},
     ({\epsilon_2},{\epsilon_3})} \right)\, =\,
  F^{AC}\left(\Gamma_{({\epsilon_1},{\epsilon_2}),
      {\epsilon_3}} \right).
 \label{equivcond}
\end{equation}
One sees that it is necessary and sufficient to
assume that $\omega\equiv 1$ or $G$ is a commutative group.
In this case, there occurs much simplification in, e.g. eqs
(\ref{theta}) and (\ref{gamma}),  and it can be shown shown that
eq.(\ref{equivcond}) is satisfied by explicitly checking for every
choice of $\epsilon_i$ ($i=1,2,3$).

{}From now on, we assume that $\omega\equiv 1$ or $G$ is a commutative group.
Then the linear transformation $F^{AC}(\Gamma)$ acquires more symmetries.
It commutes with cyclic permutations
\begin{equation}
  F^{AC}\left(\Gamma_{({\epsilon_1},{\epsilon_2},
        {\epsilon_3})} \right)\, =\,
 P_{231}^{-1}\circ
  F^{AC}\left(\Gamma_{({\epsilon_2},
{\epsilon_3},{\epsilon_1})} \right)  \circ P_{231}.
    \label{cyclic}
\end{equation}
Furthermore, it satisfies
\begin{eqnarray}
  P_{321} \circ
  \left( F^{AC}(\Gamma_{({\epsilon_1},{\epsilon_2},
        {\epsilon_3})}) \right)^{\ast}\circ P_{321}^{-1}
  \, &=&\,
   F^{AC}(\Gamma_{({-\epsilon_3},{-\epsilon_2},
        {-\epsilon_1})})   \label{dual}       \\
  & & %\hspace*{1cm}
   \, : \, V^{-\epsilon_3}\otimes V^{-\epsilon_2}\otimes
    V^{-\epsilon_1} \longrightarrow
    V^{-\epsilon_3}\otimes V^{-\epsilon_2}\otimes
    V^{-\epsilon_1} \nonumber
  \end{eqnarray}
where $\ast$ means dual.
This relation shows that
$\pi$ rotation  of the graph
$\Gamma_{({\epsilon_1},{\epsilon_2},{\epsilon_3})}$
within the plane where $\Gamma$ lies
corresponds to taking the dual map of
$F^{AC}(\Gamma_{({\epsilon_1},{\epsilon_2},{\epsilon_3})})$
(Fig.\ref{lin}(d)).

Because of the symmetries mentioned above,
there are only two independent $c$-graphs
$\Gamma_{({\epsilon_1},{\epsilon_2},{\epsilon_3})}$,
which are
for $(\epsilon_1,\epsilon_2,\epsilon_3)=(1,1,1)$ and $(-1,1,1)$.
The linear transformations they induce are explicitly given by
\begin{eqnarray}
  \lefteqn{F^{AC}(\Gamma_{({+1},{+1},{+1})})\;
   (\chi_{g_1,x_1}\otimes\chi_{g_2,x_2}\otimes\chi_{g_3,x_3}) } \nonumber\\
    & = & |G| \sum_{h\in G} \delta_{g_1g_2g_3,e}
         \theta_h(g_1,g_2)\,\theta_h(g_3{}^{-1},g_3)\,
         \theta_{g_1}(h,x_1)\,\theta_{g_2}(h,x_2)\,\theta_{g_3}(h,x_3)
          \nonumber\\
     & & \hspace*{5cm} \times
     \chi_{g_1,hx_1}\otimes\chi_{g_2,hx_2}\otimes\chi_{g_3,hx_3}
   \label{f+++}
\end{eqnarray}
and
\begin{eqnarray}
  \lefteqn{F^{AC}(\Gamma_{({-1},{+1},{+1})})\;
   (\psi_{g_1,x_1}\otimes\chi_{g_2,x_2}\otimes\chi_{g_3,x_3}) } \nonumber\\
    & = & |G| \sum_{h\in G} \delta_{g_1{}^{-1}g_2g_3,e}
       \theta_h(g_2,g_3)\,
         \theta_{g_1}(h,x_1)^{-1}\,\theta_{g_2}(h,x_2)\,\theta_{g_3}(h,x_3)
          \nonumber\\
     & & \hspace*{5cm}\times
     \psi_{g_1,hx_1}\otimes\chi_{g_2,hx_2}\otimes\chi_{g_3,hx_3}  .
    \label{f-++}
\end{eqnarray}

\begin{figure}
    \begin{center}
      \leavevmode
       \epsfxsize = 15cm
       \epsfbox{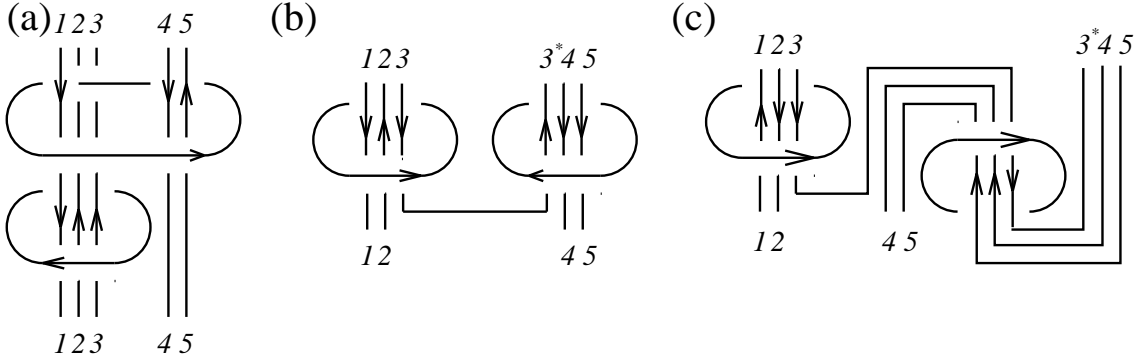}
    \end{center}
    \caption{
      (a) Connecting a bottom end and a top end.
      (b) Connecting two bottom ends.
      (c) Operation (b) is equivalent with first taking the dual
      (rotating the graph) and
      then connect a bottom end and a top end. }
   \label{connecting}
  \end{figure}

Now we `compose' these $n_2$ linear maps reflecting  the
connectivity of $3n_2$ vertical ribbons.
Let us imagine a situation such that
a top end of a building block
corresponding to the linear map
$F^{AC}(\Gamma_{(\epsilon_1,\epsilon_2,\epsilon_3)})$
is connected to a bottom
end of that corresponding to
$  F^{AC}(\Gamma_{(\epsilon_1,\epsilon_4,\epsilon_5)})$
(Fig.~\ref{connecting}(a)).
To recombine these building blocks, we consider a linear transformation
\begin{equation}
\Phi(\ei{1}\otimes\ei{2}\otimes\ei{3}\otimes\ei{4}\otimes\ei{5})
= \sum_{j,k}f^{j_1j_4j_5}_{k_1i_4i_5}f^{k_1j_2j_3}_{i_1i_2i_3}
\eip{1}\otimes\eip{2}\otimes\eip{3}\otimes\eip{4}\otimes\eip{5}
\label{topbottom}
\end{equation}
on
$V^{\epsilon_1}\otimes V^{\epsilon_2} \otimes
V^{\epsilon_3}\otimes V^{\epsilon_4} \otimes V^{\epsilon_5}$
following the spirit of the definition of the action of $F^{AC}$ on
$c$-graphs. The basis $\{e^{\epsilon}_{k}\}_{k=1,\ldots,
  |G|^2}$ span $V^{\epsilon}$ (i.e. $e^1\sim\chi, e^{-1}\sim\psi$) and
\begin{equation}
  F^{AC}(\Gamma_{(\epsilon_1,\epsilon_2,\epsilon_3)})
(\ei{a} \otimes \ei{b} \otimes \ei{c})
= \sum_j f^{j_aj_bj_c}_{i_ai_bi_c}
\eip{a} \otimes \eip{b} \otimes \eip{c}.
\end{equation}

If two building blocks are connected through the bottom
ends of their vertical ribbons (Fig.\ref{connecting}(b)),
we consider a linear map
\begin{eqnarray}
\lefteqn{\Psi(\ei{1} \otimes \ei{2} \otimes \ei{4} \otimes \ei{5})
= \sum_{j,j_3',i_3,i'_3}
f^{j_1j_2j_3}_{i_1i_2i_3} f^{j'_3j_4j_5}_{i'_3i_4i_5} \delta_{i_3i'_3}
\eip{1}\otimes \eip{2} \otimes
\eip{3}\otimes e^{-\epsilon_3}_{j'_3} \otimes \eip{4}\otimes\eip{5}}
\nonumber\\
& : & V^{\epsilon_1} \otimes V^{\epsilon_2}\otimes
V^{\epsilon_4} \otimes V^{\epsilon_5} \rightarrow
V^{\epsilon_1}\otimes V^{\epsilon_2} \otimes
V^{\epsilon_3}\otimes V^{-\epsilon_3} \otimes V^{\epsilon_4}
\otimes V^{\epsilon_5} \label{toptop}
\end{eqnarray}
by a reason we are going to explain shortly.

We continue these operations until $6n_2$ free ends of vertical
ribbons are contracted.
Finally we are led to the definition of a number $F^{AC}(L_T)$ :
\begin{equation}
  F^{AC}(L_T) =  \sum_{i_1,\ldots,i_{6n_2}}
 %f^{i_{3n_2+1},\ldots,j_{6n_2}}_{i_1,\ldots,i_{3n_2}}
 f^{i_{3n_2+1},i_{3n_2+2},i_{3n_2+3}}_{i_1,i_2,i_3}
 f^{i_{3n_2+4},i_{3n_2+5},i_{3n_2+6}}_{i_4,i_5,i_6}\cdots
 f^{i_{6n_2-2},i_{6n_2-1},i_{6n_2}}_{i_{3n_2-2},i_{3n_2-1},i_{3n_2}}
\prod_{n=1}^{3n_2} \delta_{i_{a_n},i_{b_n}},
 \label{contraction}
\end{equation}
where $(a_n,b_n)$ $(n=1,\ldots, 3n_2)$ are the pairs of ends of ribbons
that are connected in $L_T$.
We see that $F^{AC}(L_T)$ does not change when we
change the order of the tensor product in eqs.(\ref{topbottom}),(\ref{toptop}).

It might seem more natural to use
the action of $F^{AC}$ on  various graphs, e.g.
eqs.(\ref{op:cupright}) or (\ref{op:cupleft})
(eqs.(\ref{op:capright}) or (\ref{op:capleft}) )
when we connect two bottom (top) ends.
However, because we have decided which end is the top of a graph by hand,
it is expected that the final result (\ref{contraction}) is
invariant under interchanging the top end and the bottom end of
a graph (i.e. taking the dual of one of the linear maps by
eq.(\ref{dual}). See Fig.~\ref{connecting}(c).)
It is necessary to define the linear map by  eq.(\ref{toptop})
in order to have the invariance.

Furthermore, we  verify that this quantity
$F^{AC}(L_T)$
is a topological
invariant of
three-manifold  when multiplied by a simple numerical factor\footnote{
This factor has a simple interpretation in terms of the Kirby moves.
When we generate Alexander moves from Kirby moves, we neglect
trivial (disconnected) link components produced by the latter.
This factor is to compensate contributions of such link components.
}
 depending on $n_i$
and that it has a state-sum representation.
In fact, this is exactly the same quantity as the
Chung-Fukuma-Shapere invariant  $Z^{CFS}_{A}$
for $A=D^{\omega}(G)$.
We carry out the proof of the invariance
using the equivalence with the Chung-Fukuma-Shapere invariant
$Z^{CFS}_{D^\omega(G)}$.
We comment that the proof can be performed
following the idea of generating the Alexander moves from the Kirby
moves, without mentioning the Chung-Fukuma-Shapere invariant.

Before doing this we briefly explain the
Chung-Fukuma-Shapere invariant $Z^{CFS}_A(M)$.
It is defined for each involutory Hopf algebra $A$,
where involutory means that the square of the antipode operator
$S : A\rightarrow A$ is the identity~\cite{cfs,ku}.
We provide a lattice of $M$ and
decompose it into the set of polygonal faces $F=\{ f\}$
and `hinges' $H=\{ h \}$ which have a role of connecting faces.
We put an arrow on each edge of each face $f$ by a cyclic order
around a face,
which also induces arrows on hinges.
By assigning $C_{f_1f_2\cdots f_n}$ and
$\Delta^{h_1\cdots h_j '\cdots h_n}\prod_j S^{h_j}{}_{h_j'}$
to each $n$-face and $n$-hinge,
the partition function  $Z^{CFS}_{A}(M)$
is defined as~\cite{cfs,ku}
  \begin{equation}
    Z^{CFS}_{A}(M)
    \equiv
    |\dim A|^{-n_3-n_1}
    \prod_{f\in F}  C_{x_{f[1]}\cdots x_{f[n_f]}}
    \prod_{h\in H}
    \left[\Delta^{x_{h[1]}\cdots x'_{h[j]} \cdots x_{h[n_h]}}
    \prod_{j \in R_h} S^{x_{h[j]}}{}_{x'_{h[j]}} \right].
    \label{zfuk}
  \end{equation}

For $A=D^\omega(G)$, the theory is defined if
$\omega\equiv 1 $ or $G$ is commutative.
It is characterized by the data $C,\Delta$ and  $S$  given as
\begin{eqnarray}
 C_{(g_1,h_1)(g_2,h_2)\cdots(g_k,h_k)}
  &=& |G|\delta_{g_1,g_2}\delta_{g_1,g_3}\cdots\delta_{g_1,g_k}
  \delta_{h_1\cdots h_k, e} \times \\ \nonumber
  & & \times \prod_{j=1}^{k-1} \theta_{g_1}
  \left(\prod_{\ell=1}^{j}h_\ell,h_{j+1}\right)
  \qquad (\mbox{for\ } k\ge 3)\, ,\\
 \Delta^{(h_1,g_1)(h_2,g_2)\cdots (h_k,g_k)}
  & = & C_{(g_1,h_1)(g_2,h_2)\cdots (g_k,h_k)}\, ,\\
 S^{(h,y)}{}_{(g,x)}
  & = &  \delta_{gh,e}\delta_{xy,e}
  \theta_h(y^{-1},y)^{-1}
  \theta_{y^{-1}}(h,h^{-1})^{-1}.
\end{eqnarray}

By comparing the
eqs. (\ref{f+++}) and (\ref{f-++}) to
these data $C$, $\Delta$ and $S$,
they are rewritten as
\begin{eqnarray}
  \lefteqn{F^{AC}(\Gamma_{(+1,+1,+1)})\;
   (\chi_{g_1,x_1}\otimes\chi_{g_2,x_2}\otimes\chi_{g_3,x_3}) } \nonumber\\
    & &=  \Delta^{(a_1,b_1)(a_2,b_2)(a_3,b_3)}
       C^{(h_1,y_1)}{}_{(a_1,b_1)(g_1,x_1)}
       C^{(h_2,y_2)}{}_{(a_2,b_2)(g_2,x_2)}
       C^{(h_3,y_3)}{}_{(a_3,b_3)(g_3,x_3)}\nonumber\\
     & & \hspace*{5cm} \times\;
     \chi_{h_1,y_1}\otimes\chi_{h_2,y_2}\otimes\chi_{h_3,y_3}  \\
  \lefteqn{F^{AC}(\Gamma_{(-1,+1+1)})\;
   (\psi_{g_1,x_1}\otimes\chi_{g_2,x_2}\otimes\chi_{g_3,x_3}) } \nonumber\\
    & = & \Delta^{(a_1,b_1)(a_2,b_2)(a_3,b_3)} S^{(a_1',b_1')}{}_{(a_1,b_2)}
       C^{(g_1,x_1)}{}_{(a'_1,b'_1)(h_1,y_1)}
       C^{(h_2,y_2)}{}_{(a_2,b_2)(g_2,x_2)}
       C^{(h_3,y_3)}{}_{(a_3,b_3)(g_3,x_3)}\nonumber\\
         & & \hspace*{5cm}\times\;
     \psi_{h_1,y_1}\otimes\chi_{h_2,y_2}\otimes\chi_{h_3,y_3}  .
    \label{cfs-++}
\end{eqnarray}

Here we see that the number
$F^{AC}(L_T)$
defined by eq.(\ref{contraction})
is identical with the partition  function $Z^{CFS}_{A=D^\omega(G)}$
(\ref{zfuk}) for the dual lattice $T^\ast$ of $T$
up to a normalization factor, i.e.,
  \begin{equation}
    |G|^{-2(n_0+n_2)} \times F^{AC}(L_T) = Z^{CFS}_{A=D^\omega(G)}(T^{\ast}).
  \end{equation}

We started with the Altschuler-Coste
functor $F^{AC}$ which gave an invariant for framed links.
We `operated' the functor on a framed link $L_T$ associated with
the triangulation $T$ and ended up with
the Chung-Fukuma-Shapere theory for  $D^\omega(G)$.
Originally, the functor $F^{AC}$ was applied to framed links along
which surgeries were performed and Altschuler-Coste theory was obtained.
So we have two distinct topological field theories via
the lattice construction and the surgery construction
starting with a single functor $F^{AC}$.
It is known that
the Chung-Fukuma-Shapere theory for $D^\omega(G)$
is related to the Altschuler-Coste theory for $D^\omega(G)$
in the sense that the former is a tensor product of two latter
theories for $G=\Z_{2N+1}$ \cite{ah}.
Therefore in some cases the lattice and surgery
constructions produce the identical topological field theory (up to
the square).

\bigskip

In ref.\cite{bou2}
it is suggested that
a link representation of a manifold
$L_T$ and the Reshetikhin-Turaev functor $F^{RT}_{\uqsl}$
for $q=\exp (2\pi i/r)$
induces the Turaev-Viro invariant
of three-manifolds.
Roughly speaking, the idea is as follows.
Define
 \begin{eqnarray}
%   \vspace{2cm}
  F^{RT}_q(j_1,j_2,j_3) & \equiv &
  \sum_{\lambda\in I}\; [2\lambda +1]_q \;
   F^{RT}_{\uqsl}\Bigg(\;
    \begin{picture}(90,30)
     \put(40,3){\oval(70,20)[b]}
     \put(21,3){\oval(32,20)[tl]}
     \put(59,3){\oval(32,20)[tr]}
     \put(29,13){\line(1,0){7}}
     \put(44,13){\line(1,0){7}}
     \put(25,-3){\line(0,1){30}}
     \put(25,-10){\line(0,-1){15}}
     \put(40,-3){\line(0,1){30}}
     \put(40,-10){\line(0,-1){15}}
     \put(55,-3){\line(0,1){30}}
     \put(55,-10){\line(0,-1){15}}
     \put(20,-35){$j_1$}
     \put(35,-35){$j_2$}
     \put(50,-35){$j_3$}
     \put(77,-10){$\lambda$}
   \end{picture}
      \; \Bigg)
      \\
  & & \rule{0cm}{2cm}
  \hspace*{2.5cm}
   \, : \, V^{j_1}\otimes V^{j_2}\otimes V^{j_3} \rightarrow
                V^{j_1}\otimes V^{j_2}\otimes V^{j_3} \nonumber
    \label{frt123}
 \end{eqnarray}
for a ribbon $(3,3)$-graph colored by
$j\in\{0,1/2,1,\cdots , r/2-1 \}$.
The vector space $V^j$ is $2j+1$ dimensional and is spanned by
$b^j_m$ $(m=j,j-1,\cdots,-j)$~\cite{km}.
It is calculated as
 \begin{eqnarray}
  F^{RT}_q(j_1,j_2,j_3)
     (b^{j_1}_{m_1}\otimes b^{j_2}_{m_2}\otimes b^{j_3}_{m_3}) &&
            \nonumber\\  & & \hspace{-6cm} =
\delta(j_1 j_2 j_3)\;
      \sum_{n_1, n_2, n_3}
     {\btuple{j_1}{j_2}{j_3}{m_1}{m_2}{-m_3}}_q
     {\btuple{j_1}{j_2}{j_3}{n_1}{n_2}{-n_3}}_q
     \nonumber\\ & & \hspace{-6cm}
     \times [2j_3+1]_q{}^{- 1}
     q^{\frac {m_3+n_3}2} (-1)^{2 j_3+m_3 +n_3} \;
     b^{j_1}_{n_1}\otimes b^{j_2}_{n_2}\otimes b^{j_3}_{n_3}
   \label{rt123}
 \end{eqnarray}
where
$[:::]_q$ is the $q$-$3j$-symbol and
$\delta(j_1 j_2 j_3)$ takes 1 if
$j_1\le j_2+j_3, j_2\le j_1+j_3, j_3\le j_1+j_2, j_1+j_2+j_3\le r-2$
and $j_1+j_2+j_3$ is integer, and $0$ otherwise.
The claim is that the combination of four
$q$-$3j$-symbols from (\ref{rt123}) gives rise to $q$-$6j$-symbol,
which yields the Turaev-Viro invariant.

However, (\ref{rt123}) shows us that the cyclic symmetry
$F^{RT}_q(j_1,j_2,j_3)= F^{RT}_q(j_2,j_3,j_1)$
holds only in the case of $q=1$.
It means that  in the case of $q\neq 1$
there is no well-defined way of
interpreting each building block of $L_T$  as a ribbon (3,3)-graph.
It follows that we cannot construct the Turaev-Viro
theory from the Reshetikhin-Turaev functor $F^{RT}_{\uqsl}$
by this method, though there are connections
between them~\cite{kr}.

On the other hand, it can be shown that the set of
$F^{RT}_q(j_1,j_2,j_3)$
produce the Turaev-Viro invariant in the limit $q\rightarrow 1$, i.e., the
Ponzano-Regge partition function \cite{pr},
by the similar equation as (\ref{contraction})
defined by (\ref{rt123}).

%%%%%%%%%%%%%%%%%%%%%%%%%%%%%%%%%%%%%%%%%%%%%%%%%%%%%%%%%%%%%%%%%%%%%%
%\paragraph{Acknowledgments}
%%%%%%%%%%%%%%%%%%%%%%%%%%%%%%%%%%%%%%%%%%%%%%%%%%%%%%%%%%%%%%%%%%%%%%

%\bibitem{rt2}
%N.~Reshetikhin and V.~G.~Turaev,
%{\sl Invent. Math.} {\bf 103}(1991)547.

%\bibitem{ac2}
%D.~Altschuler and A.~Coste,
%{\sl J. Geom. Phys} {\bf 11}(1993)191.

%\bibitem{bou}
%D.~V.~Boulatov,
%{\sl Mod. Phys. Lett.} {\bf A7} (1992)1629.


\begin{thebibliography}{1}

\bibitem{a}
M.~Atiyah,
{\sl Publ. IHES} {\bf 68}(1989)175.

\bibitem{dw}
R. Dijkgraaf and E. Witten,
{\sl Commun. Math. Phys.}{\bf 129}(1990)393.

\bibitem{tv}
V.~G.~Turaev and O.~Y.~Viro,
{\sl Topology} {\bf 31}(1992)865.

\bibitem{rt}
N.~Yu.~Reshetikhin and V.~G.~Turaev,
{\sl Commun. Math. Phys.} {\bf 127} (1990) 127.

\bibitem{dj}
B.~Durhuus and T.~J\'{o}nsson,
{\sl J. Math. Phys.} {\bf 35}(1994)5306.

\bibitem{cfs}
S.~Chung, M.~Fukuma, and A.~Shapere,
{\sl Int. J. Mod. Phys.} {\bf A9}(1994)1305.

\bibitem{wi}
E.~Witten,
{\sl Commun. Math. Phys.} {\bf 121} (1989)351. % jones polynomial

\bibitem{tu}
V.~G.~Turaev,
{\sl C. R. Acad. Sci. Paris} {\bf 313}(1991)395.

\bibitem{ac}
D.~Altschuler and A.~Coste,
{\sl Commun. Math. Phys.} {\bf 150}(1992)83.

\bibitem{ah}
M.~Asano and S.~Higuchi,
{\sl Mod. Phys. Lett.} {\bf A9} (1994)2359.

\bibitem{bou2}
D.~V.~Boulatov,
{\sl Int. J. Mod. Phys.} {\bf A8} (1993)3139.

\bibitem{kir}
R.~Kirby,
{\sl Invent. Math.} {\bf 45}(1978)35.

\bibitem{al}
J.~W.~Alexander,
{\sl Ann. Math.\/} {\bf 31}(1930)292.

\bibitem{ku}
G.~Kuperberg,
{\sl Int. J. Math.} {\bf 2}(1991)41.

\bibitem{km}
R.~Kirby and P.~Melvin,
{\sl Invent. Math.} {\bf 105}(1991)473.

\bibitem{kr}
A.~N.~Kirillov and  N.~Yu.~Reshetikhin,
{\sl Adv. Ser. in Math. Phys., 1988, pp.285--339, ed. V.~G.~Kac}.

\bibitem{pr} G.~Ponzano and T.~Regge,
{\sl Spectroscopic and Group Theoretical Methods in Physics\/},
ed. F.~Bloch (North-Holland, Amsterdam, 1968).
\end{thebibliography}
\end{document}